\documentclass[pra,aps,twocolumn,showpacs]{revtex4}

\usepackage{amsmath}
\usepackage{bm}
\usepackage{graphicx}

\begin{document}

\title{Symmetry-breaking magnetization dynamics of spinor dipolar
Bose-Einstein condensates}

\author{Shoichi Hoshi and Hiroki Saito}

\affiliation{Department of Applied Physics and Chemistry,
The University of Electro-Communications, Tokyo 182-8585, Japan}

\date{\today}

\begin{abstract}
Symmetry-breaking magnetization dynamics of a spin-1 Bose-Einstein
condensate (BEC) due to the dipole-dipole interaction are investigated
using the mean-field and Bogoliubov theories.
When a magnetic field is applied along the symmetry axis of a
pancake-shaped BEC in the $m = 0$ hyperfine sublevel, transverse
magnetization develops breaking the chiral or axial symmetry. 
A variety of magnetization patterns are formed depending on the
strength of the applied magnetic field.
The proposed phenomena can be observed in $^{87}{\rm Rb}$ and $^{23}{\rm
Na}$ condensates.
\end{abstract}


\maketitle

\section{Introduction}

A Bose-Einstein condensate (BEC) of atoms with spin degrees of freedom
(spinor BEC) allows the study of magnetism in a quantum fluid.
There are two mechanisms of magnetization for a spinor BEC: the
ferromagnetic contact interaction and the magnetic dipole interaction
(MDI) between atoms.
Magnetization dynamics due to the ferromagnetic contact interaction have
been observed for a spin-1 $^{87}{\rm Rb}$
BEC~\cite{Chang,Sadler}.
However, magnetization dynamics due to the MDI have not been studied
yet, and this is the subject of the present paper.

A BEC of $^{52}{\rm Cr}$ atoms with a large magnetic dipole moment
(6 $\mu_{\rm B}$ with $\mu_{\rm B}$ being the Bohr magneton) has been
realized by the Stuttgart group~\cite{Gries05} and its anisotropic
behaviors originating from the anisotropy of the dipole-dipole
interaction have been
observed~\cite{Gries06,Lahaye07,Koch,Lahaye08,Metz}.
While the magnetic dipole moment of spin-1 alkali atoms ($\mu_{\rm B} /
2$) is much smaller than that of $^{52}{\rm Cr}$, it is nevertheless
predicted that a small MDI can create spin textures in a $^{87}{\rm Rb}$
BEC~\cite{Yi,Kawaguchi06_2,Kawaguchi07}.
The crystalline magnetic order observed in a $^{87}{\rm Rb}$ BEC is
considered to be caused by the MDI~\cite{Venga}.
MDI effects have been detected in $^{39}{\rm K}$~\cite{Fattori} and
$^7{\rm Li}$~\cite{Pollack} BECs using Feshbach resonance.

In this paper, we show that magnetization dynamically develops in spin-1
$^{87}{\rm Rb}$ and $^{23}{\rm Na}$ BECs due to the MDI.
We consider a situation in which a BEC in the $m = 0$ magnetic sublevel
is confined in an axisymmetric pancake-shaped trap and a magnetic field
is applied along the symmetry axis.
We numerically solve the time-dependent nonlocal Gross-Pitaevskii (GP)
equation including the MDI and show that magnetization develops in the
direction perpendicular to the magnetic field due to the MDI.
These magnetization dynamics break the chiral or axial symmetry
spontaneously.
We find that various magnetization patterns emerge depending on the
strength of the magnetic field.
For spin-1 $^{87}{\rm Rb}$, we can suppress magnetization due to the
ferromagnetic contact interaction using the microwave-induced quadratic
Zeeman effect~\cite{Leslie}, and the pure MDI effect can thus be
observed.
We perform a Bogoliubov analysis and show that the magnetization is
triggered by the dynamical instability.
We also employ the variational method with the Gaussian approximation to
explain the numerical results.

This paper is organized as follows.
Section~\ref{s:formulation} formulates the mean-field and Bogoliubov
theories to study the present system.
Section~\ref{s:dynamics} numerically studies the Bogoliubov spectra and
demonstrates the magnetization dynamics.
Section~\ref{s:Gauss} analyzes the phenomena using the variational
method.
Section~\ref{s:conclusion} gives conclusions to the study.

\section{Formulation of the problem}
\label{s:formulation}

\subsection{Mean-field theory}

We consider a system of spin-1 bosonic atoms with mass $M$ and magnetic
dipole moment $\mu_{\rm B} / 2$ confined in an axisymmetric harmonic
potential $V(\bm{r}) = M (\omega_\perp^2 r_\perp^2 + \omega_z^2 z^2) /
2$ with $r_\perp = (x^2 + y^2)^{1/2}$.
A uniform magnetic field $B_z$ is applied in the $z$ direction and the
linear Zeeman energy is given by $-\mu_{\rm B} B_z m/ 2$ for magnetic
sublevels $m = 0, \pm 1$.
We neglect the magnetic quadratic Zeeman effect, since the strength of
the magnetic field considered here is $B_z < 10$ mG.
Instead, we assume that the microwave-induced quadratic Zeeman
effect~\cite{Leslie} lifts the energy of the $m = \pm 1$ states by $Q$.

We employ the mean-field approximation and the condensate is described
by the macroscopic wave function $\psi_m$ with magnetic sublevel $m$,
which satisfies the normalization condition $\sum_m \int |\psi_m|^2
d\bm{r} = N$, with $N$ being the total number of atoms.
The nonlocal GP equations including the MDI are given by
\begin{subequations}
\label{GP}
\begin{eqnarray}
& & i \hbar \frac{\partial \psi_0}{\partial t} = \left(
-\frac{\hbar^2}{2M} \nabla^2 + V + g_0 \rho \right) \psi_0
\nonumber \\
& & + \frac{g_1}{\sqrt{2}} \left( F_+ \psi_1 + F_- \psi_{-1} \right) -
\frac{\mu_{\rm B}}{2} \bm{B}_{\rm d} \cdot \sum_m (\bm{f})_{0m} \psi_m,
\nonumber \\
\end{eqnarray}
\begin{eqnarray}
& & i \hbar \frac{\partial \psi_{\pm 1}}{\partial t} = \left(
-\frac{\hbar^2}{2M} \nabla^2 + V \mp P + Q + g_0 \rho \right)
\psi_{\pm 1}
\nonumber \\
& & + g_1 \left( \frac{1}{\sqrt{2}} F_\mp \psi_0 \pm F_z
\psi_{\pm 1} \right)
- \frac{\mu_{\rm B}}{2} \bm{B}_{\rm d} \cdot \sum_m (\bm{f})_{\pm
1m} \psi_m,
\nonumber \\
\label{GPb}
\end{eqnarray}
\end{subequations}
where $\rho = \sum_m |\psi_m|^2$, $\bm{F} = \sum_{mm'} \psi_m^*
(\bm{f})_{mm'} \psi_{m'}$, $\bm{f}$ is the vector of the spin-1
matrices, $F_{\pm} = F_x \pm i F_y$, and
\begin{equation}
P = \frac{\mu_{\rm B}}{2} B_z.
\end{equation}
The spin-independent and spin-dependent contact-interaction parameters
$g_0$ and $g_1$ have the forms
\begin{equation}
g_0 = \frac{4\pi \hbar^2}{M} \frac{a_0 + 2 a_2}{3}, \qquad
g_1 = \frac{4\pi \hbar^2}{M} \frac{a_2 - a_0}{3},
\end{equation}
where $a_S$ is the {\it s}-wave scattering length for colliding atoms
with total spin $S$.
The MDI produces an effective magnetic field:
\begin{equation} \label{Bd}
\bm{B}_{\rm d}(\bm{r}) = -\frac{\mu_0}{4\pi} \frac{\mu_{\rm B}}{2} \int
d\bm{r}' \frac{\bm{F}(\bm{r}') - 3 [\bm{F}(\bm{r}') \cdot \bm{e}]
\bm{e}} {|\bm{r} - \bm{r}'|^3},
\end{equation}
where $\mu_0$ is the magnetic permeability of vacuum and $\bm{e} =
(\bm{r} - \bm{r}') / |\bm{r} - \bm{r}'|$.
Equation~(\ref{GP}) is numerically solved in Fourier space for the
kinetic term and in real space for the other terms using a fast
Fourier transform (FFT).
The FFT is also used to calculate the convolution integral in
Eq.~(\ref{Bd}).

In the present paper, we consider an initial state in which all the
atoms are in the $m = 0$ sublevel.
In order to simulate this situation, we prepare the ground state of
$\psi_0$ with $\psi_{\pm 1} = 0$ by the imaginary-time evolution of
Eq.~(\ref{GP}).
Small noise (a complex random number on each mesh) is then applied to
the initial state of $\psi_{\pm 1}$ to break the symmetry and trigger
magnetization due to dynamical instability.
The small noise corresponds to quantum fluctuation, thermal atoms, and
residual atoms in an experiment.
We note that if the initial state of $m = \pm 1$ is $\psi_{\pm 1} = 0$,
the right-hand side of Eq.~(\ref{GPb}) vanishes and $\psi_{\pm 1}$ never
develops within the mean-field theory.
Magnetization thus occurs if small noise in the $m = \pm 1$ state is
exponentially amplified by dynamical instabilities.

\subsection{Bogoliubov analysis}

We investigate the Bogoliubov excitation spectrum of magnons for the
stationary state of $\psi_0$.
Assuming that $\psi_{\pm 1}$ is small and neglecting the second and
higher orders of $\psi_{\pm 1}$ in Eq.~(\ref{GP}), we obtain
\begin{eqnarray} \label{bogoeq}
& & i \hbar \frac{\partial \psi_{\pm 1}}{\partial t} = \left(
-\frac{\hbar^2}{2M} \nabla^2 + V \mp P + Q + g_0 |\psi_0|^2
\right) \psi_{\pm 1} 
\nonumber \\
& & + g_1 \left( |\psi_0|^2 \psi_{\pm 1} +
\psi_0^2 \psi_{\mp 1}^* \right)
- \frac{g_{\rm d}}{2 \sqrt{2}} \int d\bm{r}' \frac{1}{|\bm{r} -
\bm{r}'|^3} 
\nonumber \\
& & \times \left[ (1 - 3 e_z^2) F_{\mp}(\bm{r}') + 3 e_{\mp}^2
F_{\pm}(\bm{r}') \right] \psi_0(\bm{r}),
\nonumber \\
\end{eqnarray}
where $g_{\rm d} = \mu_0 \mu_{\rm B}^2 / (16 \pi)$ and $e_{\pm} = e_x
\pm i e_y$.
Using the mode functions $u_{\pm 1}$ and $v_{\pm 1}$, we write a
single-mode excitation of $\psi_{\pm 1}$ as
\begin{eqnarray} \label{psibogo}
\psi_{\pm 1}(\bm{r}, t) & = & e^{-i \mu t / \hbar} \Bigl[
u_{\pm 1}(r_\perp, z) e^{i (L \pm 1 - 1) \phi} e^{-i \omega t}
\nonumber \\
& & + v_{\pm 1}^*(r_\perp, z) e^{-i (L \mp 1 - 1) \phi} e^{i \omega t}
\Bigr], 
\end{eqnarray}
where $\phi = {\rm arg}(x + i y)$, $L$ is an integer, and $\mu$ is the
chemical potential,
\begin{equation} \label{mu}
\mu = \frac{1}{N} \int d\bm{r} \psi_0^* \left( -\frac{\hbar^2}{2M}
\nabla^2 + V + g_0 |\psi_0|^2 \right) \psi_0.
\end{equation}
Each $\psi_{\pm 1}$ in Eq.~(\ref{psibogo}) has $2 (L - 1)$-fold
symmetry around the $z$ axis.
Substituting Eq.~(\ref{psibogo}) and $\psi_0 = |\psi_0| \exp(-i \mu t /
\hbar)$ into Eq.~(\ref{bogoeq}) yields the closed form of the nonlocal
Bogoliubov-de Gennes equations,
\begin{widetext}
\begin{subequations} \label{BdG}
\begin{eqnarray}
& & \left\{ -\frac{\hbar^2}{2M} \left[ \nabla_{\perp z}^2 - \frac{(L
\pm 1 - 1)^2}{r_\perp^2} \right] + V \mp P + Q + g_0 |\psi_0|^2 -
\mu \right\} u_{\pm 1} + g_1 |\psi_0|^2 \left( u_{\pm 1} + v_{\mp
1} \right) 
\nonumber \\
& & - \frac{g_{\rm d}}{2} \int d\bm{r}'
\frac{|\psi_0(\bm{r}')|}{|\bm{r} - \bm{r}'|^3} \left\{ (1 - 3 e_z^2)
\left[ u_{\pm 1}(\bm{r}') + v_{\mp 
1}(\bm{r}') \right] + 3 e_\mp^2 \left[ u_{\mp 1}(\bm{r}') + v_{\pm
1}(\bm{r}') \right] \right\} |\psi_0(\bm{r})|
= \hbar \omega u_{\pm 1},
\\
& & \left\{ -\frac{\hbar^2}{2M} \left[ \nabla_{\perp z}^2 - \frac{(L
\mp 1 - 1)^2}{r_\perp^2} \right] + V \mp P + Q + g_0 |\psi_0|^2 -
\mu \right\} v_{\pm 1} + g_1 |\psi_0|^2 \left( u_{\mp 1} + v_{\pm
1} \right)
\nonumber \\
& & - \frac{g_{\rm d}}{2} \int d\bm{r}' \frac{|\psi_0(\bm{r}')|}{|\bm{r}
- \bm{r}'|^3} \left\{ (1 - 3 e_z^2) \left[ u_{\mp 1}(\bm{r}') + v_{\pm
1}(\bm{r}') \right] + 3 e_\pm^2 \left[ u_{\pm 1}(\bm{r}') + v_{\mp
1}(\bm{r}') \right] \right\} |\psi_0(\bm{r})|
= -\hbar \omega v_{\pm 1},
\end{eqnarray}
\end{subequations}
\end{widetext}
where $\nabla_{\perp z}^2 = \partial_{r_\perp}^2 + r_\perp^{-1}
\partial_{r_\perp} + \partial_z^2$.
We numerically diagonalize Eq.~(\ref{BdG}) by expanding $u_{\pm 1}$ and
$v_{\pm 1}$ with orthogonal bases, e.g., the harmonic-oscillator
eigenfunctions.
If complex frequencies $\omega$ emerge, the stationary state $\psi_0$
becomes dynamically unstable against excitations of magnons.

\section{Magnetization dynamics and Bogoliubov spectra for alkali
atoms}
\label{s:dynamics}

\subsection{Spin-1 rubidium 87}
\label{s:Rb}

We first consider a spin-1 $^{87}{\rm Rb}$ BEC.
The scattering lengths are $a_0 = 101.8 a_{\rm B}$ and $a_2 = 100.4
a_{\rm B}$~\cite{Kempen} with $a_{\rm B}$ being the Bohr radius.
The spin-dependent contact-interaction parameter $g_1$ is
therefore negative and the ground state is ferromagnetic~\cite{Chang}.
In order to distinguish magnetization by the MDI from that by the
ferromagnetic contact interaction, we apply microwave radiation to lift
the energy of the $m = \pm 1$ states by $Q$, which must be much larger
than the ferromagnetic interaction energy $|g_1| \rho$.
The magnetization due to the ferromagnetic contact interaction is thus
suppressed and the pure effect of the MDI can be observed.

\begin{figure}[t]
\includegraphics[width=8.5cm]{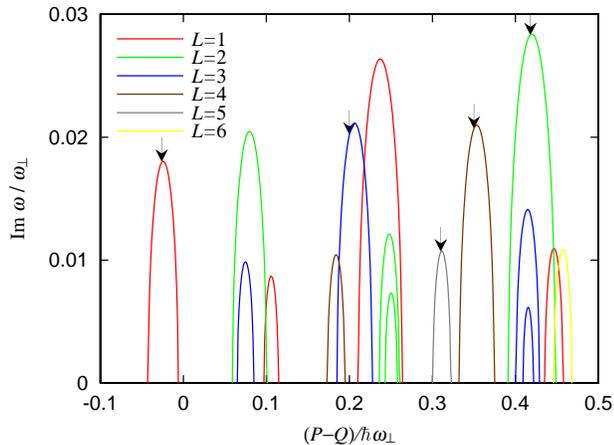}
\caption{
(color) Imaginary part of Bogoliubov frequency $\omega$ for magnon
excitation from the $m = 0$ state of $^{87}{\rm Rb}$ atoms as a function
of $P - Q$ with $Q = 50 \hbar \omega_\perp$.
The excitation mode has $2 (L - 1)$-fold symmetry around the $z$ axis,
where $L$ is defined in Eq.~(\ref{psibogo}).
The number of atoms is $N = 10^5$ and the radial and axial trap
frequencies are $\omega_\perp = 2 \pi \times 100$ Hz and $\omega_z = 2
\pi \times 400$ Hz.
The values of $P$ indicated by the arrows are used in Figs.~\ref{f:evrb}
and \ref{f:magrb}.
}
\label{f:bogorb}
\end{figure}
To investigate the dynamical instability against magnetization, we
numerically diagonalize Eq.~(\ref{BdG}).
Figure~\ref{f:bogorb} shows the imaginary part of the Bogoliubov
frequencies $\omega$ as a function of the applied magnetic field, where
$N = 10^5$ atoms are confined in a pancake-shaped trap with
$\omega_\perp = 2 \pi \times 100$ Hz and $\omega_z = 2 \pi \times 400$
Hz.
The microwave-induced quadratic Zeeman energy $Q$ is chosen to be $50
\hbar \omega_\perp$, which is sufficient to suppress magnetization by
the ferromagnetic contact interaction.
In fact, we have confirmed that the Bogoliubov spectrum is always real
if the MDI is absent, $g_{\rm d} = 0$, for this value of $Q$.
The linear Zeeman energy $P = 50 \hbar \omega_\perp$ corresponds to
$B_z \simeq 7.15$ mG.
From Fig.~\ref{f:bogorb}, we find that magnon modes with various
rotational symmetries (various $L$) become dynamically unstable,
depending on the linear Zeeman energy $P$.
There is no imaginary part for $p \equiv (P - Q) / (\hbar \omega_\perp)
< -0.1$, while many peaks in the imaginary part exist for $p > 0.5$.

\begin{figure}[t]
\includegraphics[width=8.5cm]{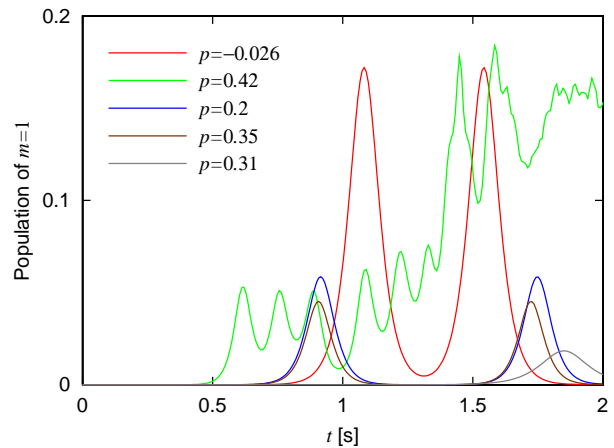}
\caption{
(color) Time evolution of the population of the $m = 1$ state $\int
|\psi_1|^2 d\bm{r} / N$ for the values of $p \equiv (P - Q) / (\hbar
\omega_\perp)$ indicated by the arrows in Fig.~\ref{f:bogorb}.
The parameters are the same as those in Fig.~\ref{f:bogorb}.
The populations of the $m = -1$ state are too small to be discerned at
this scale of the ordinate.
}
\label{f:evrb}
\end{figure}
Figure~\ref{f:evrb} shows time evolution of $\int |\psi_1|^2 d\bm{r} /
N$ for the values of $P$ indicated by the arrows in
Fig.~\ref{f:bogorb}.
The transition from the $m = 0$ state to the $m = 1$ state occurs due to
the dynamical instability shown in Fig.~\ref{f:bogorb}.
From Fig.~\ref{f:evrb}, we find that the transition occurs periodically
except for $p = 0.42$ (green line).
The complicated behavior for $p = 0.42$ originates from the fact that
the dynamically unstable modes are not only $L = 2$ but also $L = 3$
(see Fig.~\ref{f:bogorb}).
We note that the transition to the $m = -1$ state is negligibly small
and the total spin in the $z$ direction $\int (|\psi_1|^2 -
|\psi_{-1}|^2) d\bm{r}$ is not conserved, indicating that the transition
is not due to the spin-exchange contact interaction but due to the MDI.
Since the $z$ component of the total angular momentum must be conserved,
the system acquires orbital angular momentum.
The transfer of the spin angular momentum to the orbital angular
momentum in a spinor dipolar BEC also occurs in the Einstein-de Haas
effect~\cite{Kawaguchi06,Santos,Gaw}.

\begin{figure}[htbp]
\includegraphics[width=7.3cm]{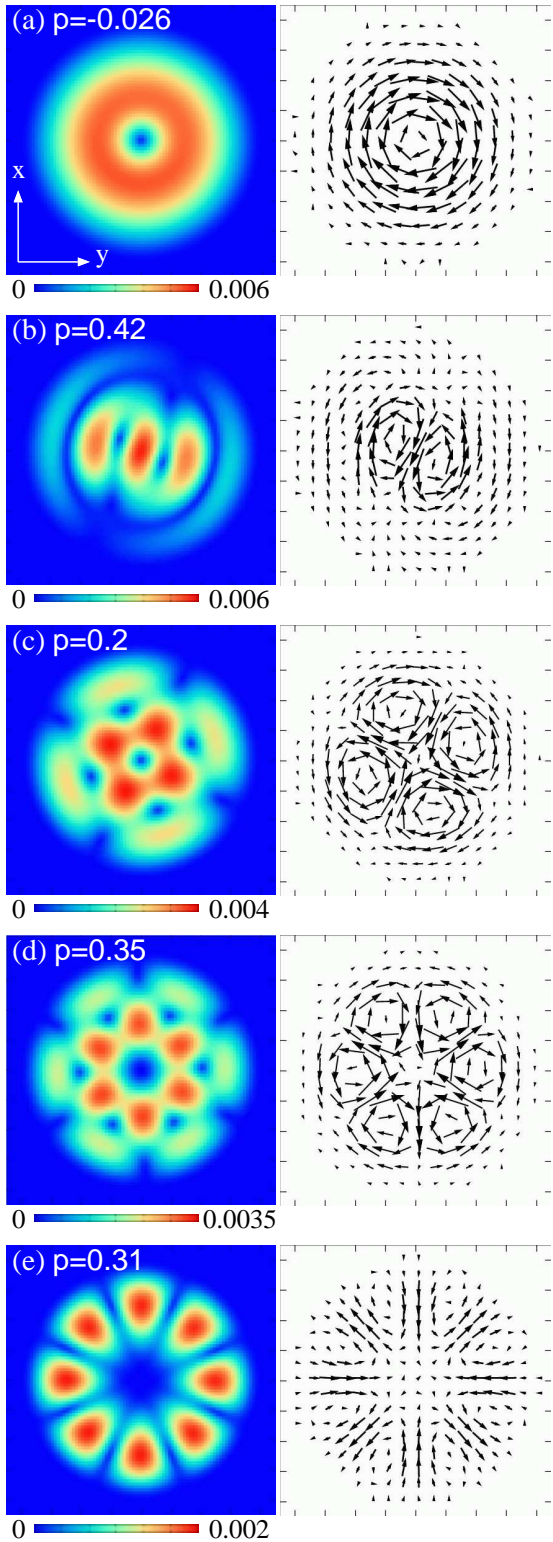}
\caption{
(color) Magnitude of the integrated transverse magnetization $|\int F_+
dz|$ (right panels) and its direction arg $(\int F_+ dz)$ (left panels)
at the first peaks of the curves in Fig.~\ref{f:evrb}.
The values of $p \equiv (P - Q) / (\hbar \omega_\perp)$ used are
indicated by the arrows in Fig.~\ref{f:bogorb}.
The unit of $\int F_+ dz$ is $N M \omega_\perp / \hbar$.
The length of the vector is proportional to $|\int F_+ dz|$.
The field of view is $9.7 \times 9.7$ $\mu {\rm m}$.
}
\label{f:magrb}
\end{figure}
Figure~\ref{f:magrb} shows transverse magnetization at the times of the
first peaks of $\int |\psi_1|^2 d\bm{r}$ (the first peaks of the lines
in Fig.~\ref{f:evrb}) for the linear Zeeman energies $P$ indicated by
the arrows in Fig.~\ref{f:bogorb}.
A variety of magnetization patterns with $2 (L - 1)$-fold symmetry
emerge depending on the strength of the applied magnetic field.
The closure structure of the magnetization in Fig.~\ref{f:evrb} (a) is
an energetically favorable structure for the MDI energy.
The directions of the magnetization vectors in the closure structure
have clockwise and counterclockwise symmetry, and 
therefore the spin-vortex generation in Fig.~\ref{f:magrb} (a) breaks
the chiral symmetry.
The $m = 1$ component of these spin vortices is $\psi_1 \propto e^{-i
\phi}$.
This situation is different from the spin-vortex generation by the
ferromagnetic contact interaction, in which polar-core vortices of
$\psi_{\pm 1} \propto e^{\pm i \phi}$ and $\psi_{\pm 1} \propto e^{\mp i
\phi}$ emerge with an equal probability~\cite{SaitoL}.
The closure structures are also seen in Figs.~\ref{f:magrb}
(b)-\ref{f:magrb} (e).
The magnetization in Figs.~\ref{f:magrb} (b)-\ref{f:magrb} (e) caused by
the dynamical instability with $L \geq 1$ exhibits a variety of
patterns, breaking the axisymmetry of the system.

\subsection{Spin-1 sodium 23}

\begin{figure}[t]
\includegraphics[width=8.5cm]{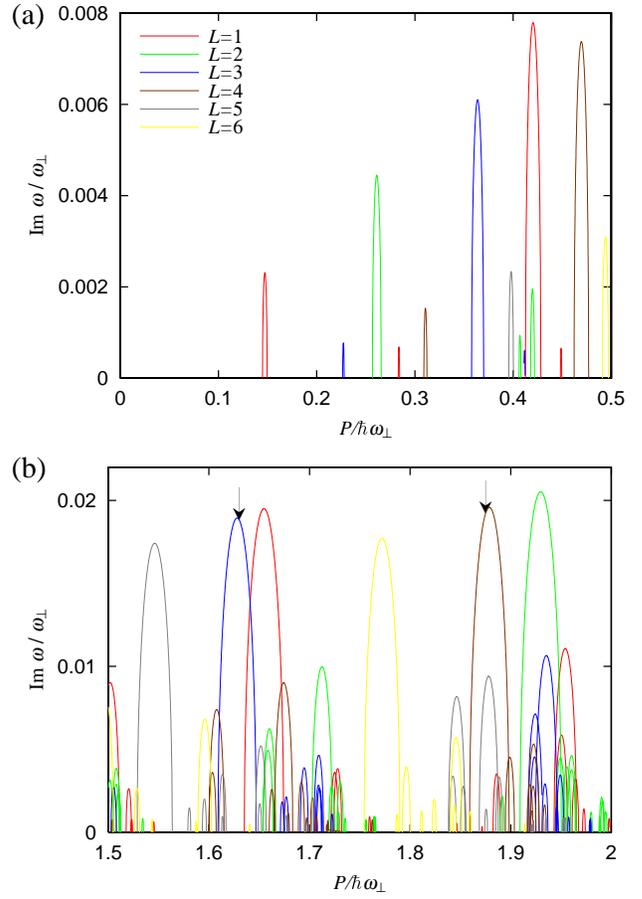}
\caption{
(color) Imaginary part of the Bogoliubov frequency $\omega$ for magnon
excitation from the $m = 0$ state of $^{23}{\rm Na}$ atoms as a function
of the linear Zeeman energy $P$.
The range of $P$ is (a) $0 \leq P / (\hbar \omega_\perp) \leq 0.5$ and
(b) $1.5 \leq P / (\hbar \omega_\perp) \leq 2$.
The excitation mode has $2 (L - 1)$-fold symmetry around the $z$ axis,
where $L$ is defined in Eq.~(\ref{psibogo}).
The number of atoms is $N = 10^6$ and the trap frequencies are the same
as those in Fig.~\ref{f:bogorb}.
The values of $P$ indicated by the arrows in (b) are used in
Fig.~\ref{f:magna}.
}
\label{f:bogona}
\end{figure}
Next we consider a spin-1 $^{23}{\rm Na}$ BEC.
The scattering lengths are given by $(a_0 + 2 a_2) / 3 = 53.4 a_{\rm
B}$~\cite{Crub} and $a_2 - a_0 = 2.47 a_{\rm B}$~\cite{Black}.
The spin-dependent contact-interaction parameter $g_1$ is then
positive and the polar state ($m = 0$) is energetically favorable.
Spontaneous magnetization due to the contact interaction is therefore
suppressed and the microwave-induced Zeeman effect is unnecessary ($Q =
0$).
The number of atoms is assumed to be $N = 10^6$ and the trap frequencies
are the same as those in Sec.~\ref{s:Rb}.

Figure~\ref{f:bogona} shows the imaginary part of the Bogoliubov
frequency obtained by numerically diagonalizing Eq.~(\ref{BdG}).
Compared with the case of $^{87}{\rm Rb}$ in Fig.~\ref{f:bogorb}, the
width and height of the peaks are small for $0 \leq P / (\hbar
\omega_\perp) \leq 0.5$ [Fig.~\ref{f:bogona} (a)].
The width and height of the main peaks gradually increase and saturate
for $P / (\hbar \omega_\perp) \sim 2$ [Fig.~\ref{f:bogona} (b)].
For $P < 0$, there is no imaginary part.

\begin{figure}[tb]
\includegraphics[width=8.0cm]{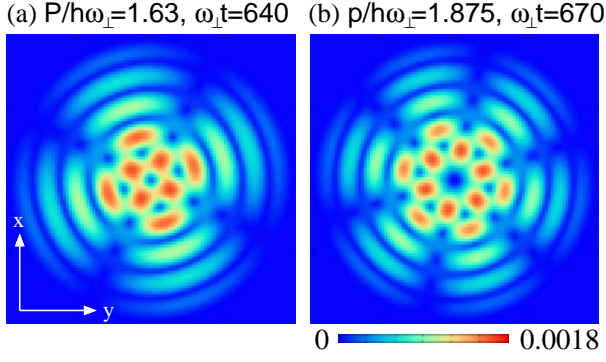}
\caption{
(color) Integrated transverse magnetization $|\int F_+ dz|$ at the time
when $\int |\psi_1|^2 d\bm{r}$ becomes the first maximum in time
evolution for the values of $P$ indicated by the arrows in
Fig.~\ref{f:bogona} (b).
The unit of $|\int F_+ dz|$ is $N M \omega_\perp / \hbar$.
The field of view is $21 \times 21$ $\mu {\rm m}$.
}
\label{f:magna}
\end{figure}
We numerically solve the GP equation for the values of $P$ indicated by
the arrows in Fig.~\ref{f:bogona} (b).
The initial state is prepared by the same method as for $^{87}{\rm
Rb}$.
Figure~\ref{f:magna} shows the integrated transverse magnetization
$|\int F_+ dz|$ at the time of the first peak of $\int |\psi_1|^2 d{\bm
r}$ in the time evolution.
Many radial nodes in the magnetization patterns are evident, since the
values of $P$ correspond to the higher-order peaks in
Fig.~\ref{f:bogona} (b).
The population of the $m = 1$ state, $\int |\psi_1|^2 d\bm{r} / N$, is 0.01
in Fig.~\ref{f:magna} (a) and 0.05 in Figure~\ref{f:magna} (b).
The population of the $m = -1$ state is very small $\sim 10^{-4}$.

\section{Gaussian variational analysis}
\label{s:Gauss}

To qualitatively examine the Bogoliubov spectra obtained in
Sec.~\ref{s:dynamics}, we perform Gaussian variational analysis.
The variational wave function for the $m = 0$ state has the form
\begin{equation} \label{var0}
\psi_0 = \frac{\sqrt{N}}{\pi^{3/4} d_\perp d_z^{1/2}}
\exp\left(-\frac{r_\perp^2}{2d_\perp^2} - \frac{z^2}{2d_z^2} - i
\frac{\mu}{\hbar} t \right),
\end{equation}
where $d_\perp$ and $d_z$ are the variational parameters characterizing
the size of the condensate in the radial and axial directions.
Substituting Eq.~(\ref{var0}) into Eq.~(\ref{mu}) and the mean-field
energy
\begin{equation}
E = \int d\bm{r} \psi_0^* \left( -\frac{\hbar^2}{2M} \nabla^2 + V +
\frac{g_0}{2} |\psi_0|^2 \right) \psi_0,
\end{equation}
we obtain
\begin{eqnarray}
\frac{\mu}{\hbar \omega_\perp} & = & \frac{1}{2} \left( \frac{1}{\tilde
d_\perp^2} + \tilde d_\perp^2 \right) + \frac{1}{4} \left(
\frac{1}{\tilde d_z^2} + \lambda^2 \tilde d_z^2 \right) + \frac{\tilde
g_0}{\tilde d_\perp^2 \tilde d_z}, \nonumber \\
\label{varmu}
\\
\frac{E}{N\hbar \omega_\perp} & = & \frac{1}{2} \left( \frac{1}{\tilde
d_\perp^2} + \tilde d_\perp^2 \right) + \frac{1}{4} \left(
\frac{1}{\tilde d_z^2} + \lambda^2 \tilde d_z^2 \right) + \frac{\tilde
g_0}{2 \tilde d_\perp^2 \tilde d_z}, \nonumber \\
\label{E}
\end{eqnarray}
where $\lambda = \omega_z / \omega_\perp$, $\tilde d_\perp = d_\perp /
a_\perp$, $\tilde d_z = d_z / a_\perp$, and $\tilde g_0 = g_0 N /
[(2\pi)^{3/2} \hbar \omega_\perp a_\perp^3]$ with $a_\perp =
[\hbar / (M \omega_\perp)]^{1/2}$.
The variational parameters $\tilde d_\perp$ and $\tilde d_z$ are
determined so as to minimize Eq.~(\ref{E}).

For simplicity, we restrict the magnon excitation to the form,
\begin{equation} \label{var1}
\psi_{\pm 1}(\bm{r}, t) = e^{-i \mu t / \hbar} \left[ \alpha_{\pm 1}
e^{-i \omega t} \chi_{\pm}(\bm{r}) + \beta_{\pm 1}^* e^{i \omega t}
\chi_{\mp}(\bm{r}) \right] ,
\end{equation}
with
\begin{equation}
\chi_\pm(\bm{r}) = \frac{e^{\pm i\phi} r_\perp}{\pi^{3/4} d_\perp^2
d_z^{1/2}} \exp\left( -\frac{r_\perp^2}{2d_\perp^2} -
\frac{r_z^2}{2d_z^2} \right),
\end{equation}
which corresponds to the lowest mode of $L = 1$ in Eq.~(\ref{psibogo}).
Substitution of Eqs.~(\ref{var0}), (\ref{varmu}), and (\ref{var1}) into
Eq.~(\ref{BdG}) yields
\begin{subequations} \label{varBdG}
\begin{eqnarray}
& & (\Lambda \mp \tilde P) \alpha_{\pm 1} + (G + D_1) (\alpha_{\pm
 1} + \beta_{\mp 1}) + D_2 (\alpha_{\mp 1} + \beta_{\pm 1})
\nonumber \\
& & = \tilde \omega \alpha_{\pm 1}, \\
& & (\Lambda \mp \tilde P) \beta_{\pm 1} + (G + D_1) (\alpha_{\mp
1} + \beta_{\pm 1}) + D_2 (\alpha_{\pm 1} + \beta_{\mp 1})
\nonumber \\
& & = -\tilde \omega \beta_{\pm 1},
\end{eqnarray}
\end{subequations}
where $\tilde P = P / (\hbar \omega_\perp)$, $\tilde \omega = \omega /
\omega_\perp$, and
\begin{eqnarray}
\Lambda & = & \frac{1}{2} \left( \frac{1}{\tilde d_\perp^2} + \tilde
d_\perp^2 \right) + \frac{Q}{\hbar \omega_\perp} - \frac{\tilde g_0}{2
\tilde d_\perp^2 \tilde d_z}, \\
G & = & \frac{N g_1}{2 (2\pi)^{3/2} \hbar \omega_\perp a_\perp^3
\tilde d_\perp^2 \tilde d_z}, \\
D_1 & = & \frac{\tilde g_{\rm d}}{2 \tilde d_\perp^2 \tilde d_z (\tilde
d_\perp^2 - \tilde d_z^2)^{5/2}}
\nonumber \\
& & \times \Biggl[ (\tilde d_\perp^2 - \tilde d_z^2)^{1/2} (-4 \tilde
d_\perp^4 - 7 \tilde d_\perp^2 \tilde d_z^2 + 2 \tilde d_z^4)
\nonumber \\
& & + 9 \tilde d_\perp^4
\tilde d_z \cot^{-1} \frac{\tilde d_z}{(\tilde d_\perp^2 - \tilde
d_z^2)^{1/2}} \Biggr], \\
D_2 & = & \frac{3 \tilde g_{\rm d}}{2 \tilde d_\perp^2 (\tilde
d_\perp^2 - \tilde d_z^2)^{5/2}}
\Biggl[ \tilde d_z (\tilde d_\perp^2 - \tilde d_z^2)^{1/2} (-5 \tilde
d_\perp^2 + 2 \tilde d_z^2) 
\nonumber \\
& & + 3 \tilde d_\perp^4 \cot^{-1} \frac{\tilde d_z}{(\tilde d_\perp^2 -
 \tilde d_z^2)^{1/2}} \Biggr],
\end{eqnarray}
with $\tilde g_{\rm d} = g_{\rm d} N / [6 (2\pi)^{1/2} \hbar
\omega_\perp a_\perp^3]$.
Diagonalizing Eq.~(\ref{varBdG}), we obtain the excitation frequency as
\begin{eqnarray} \label{varomega}
\tilde \omega^2 & = & \tilde P^2 + \Lambda^2 + 2 (G + D_1) \Lambda
\nonumber \\
& & \pm 2 \sqrt{[\Lambda^2 + 2 (G + D_1) \Lambda] \tilde P^2 + \Lambda^2
D_2^2}.
\end{eqnarray}

For the parameters of $^{87}{\rm Rb}$ in Fig.~\ref{f:bogorb}, $\tilde
\omega$ in Eq.~(\ref{varomega}) becomes imaginary between $\tilde P
\simeq 49.9$ and $50.01$ and the maximum value of Im $\omega$ is $\simeq
0.01$.
For the parameters of $^{23}{\rm Na}$ in Fig.~\ref{f:bogona}, $\tilde
\omega$ becomes imaginary between $\tilde P \simeq 0.125$ and $0.13$ and
the maximum value of Im $\omega$ is $\simeq 0.002$.
These results are in qualitative agreement with the first peaks of $L =
1$ in Figs.~\ref{f:bogorb} and \ref{f:bogona}.
The differences between the variational and numerical results come from
the forms of the variational wave functions assumed in Eqs.~(\ref{var0})
and (\ref{var1}); more appropriate variational functions are needed
for quantitative explanation of the numerical results.

\section{Conclusions}
\label{s:conclusion}

In conclusion, we have studied the magnetization dynamics caused by the
MDI in a spin-1 BEC in the $m = 0$ hyperfine state prepared in a
pancake-shaped trap and a magnetic field applied in the axial
direction.
We found that transverse magnetization develops due to the MDI breaking
the chiral or axial symmetry, and a variety of magnetization patterns
appear depending on the strength of the applied magnetic field.
We showed that these phenomena occur in spin-1 $^{87}{\rm Rb}$ and
$^{23}{\rm Na}$ BECs.
We also performed Bogoliubov analysis and found that the initial
fluctuations in the magnetization are exponentially amplified by the
dynamical instability.
A Gaussian variational analysis provided a qualitative explanation of
the results.

Our study has shown that magnetization due to the MDI strongly depends
on the shape of the system.
Magnetization dynamics for various trapping potentials including
cigar-shaped and lattice potentials merit further study.

\begin{acknowledgments}  
This work was supported by the Ministry of Education, Culture, Sports,
Science and Technology of Japan (Grants-in-Aid for Scientific Research,
No.\ 17071005 and No.\ 20540388).
\end{acknowledgments}

\end{document}